\documentclass[useAMS,usenatbib]{mn2e}
\usepackage{amsmath}
\usepackage{amssymb}
\usepackage{bm}
\usepackage{graphicx}

\title[Quenching rates for formaldehyde]{Rotational quenching of H$_2$CO by molecular hydrogen: cross-sections, rates, 
pressure broadening.}

\author[L. Wiesenfeld and A. Faure]{ %
L. Wiesenfeld $^{1}$\thanks{E-mail :  laurent.wiesenfeld@obs.ujf-grenoble.fr}, A. Faure$^{1}$\\
$^1$UJF-Grenoble 1/CNRS-INSU, Institut de Plan\'etologie et d'Astrophysique de Grenoble (IPAG) UMR 5274, Grenoble 
F-38041, France}

\date{Accepted xxx. Received xxxx; in original form \today}

\pagerange{\pageref{firstpage}--\pageref{lastpage}} \pubyear{xxxx}

\begin{document}\maketitle
\begin{abstract}

We compute the rotational quenching rates of the first 81 rotational
levels of ortho- and para-H$_2$CO in collision with ortho- and
para-H$_2$, for a temperature range of 10-300~K. We make use of the
quantum close-coupling and coupled-states scattering methods combined
with the high accuracy potential energy surface of
\citet{troscompt}. Rates are significantly different from the scaled
rates of H$_2$CO in collision with He; consequently, critical
densities are noticeably lower. We compare a full close-coupling
computation of pressure broadening cross sections with experimental
data and show that our results are compatible with the low temperature
measurements of \citet{delucia}, for a spin temperature of H$_2$
around 50~K.

\end{abstract}

\begin{keywords}
Astrochemistry, molecular processes, molecular data, ISM: molecules
\end{keywords}

\section{Introduction}

In order to relate quantitatively the observed rotational spectra and
the molecular abundances, knowledge of the relative importance of
photon induced transitions and collision induced transitions is
imperative. While the Einstein $A$ and $B$ coefficients can be
obtained experimentally (see e.g. the JPL and CDMS databases
\citet{jpl,cdms}), the rates of molecular collisional
excitation/quenching need to be calculated with help of precise
microscopic frameworks. Many molecular quenching rates have been put
forward in the last 40 years, either for molecular collisions with He
or for collisions with H, electrons and H$_2$. In many types of
interstellar regions, especially so when molecular complexity is
present, the main collider is molecular hydrogen, H$_2$. Recently, a
renewed large effort has been devoted to compute the rotational
quenching rates of molecules by H$_2$~(\textbf{see references in the
  review by \citet{vandertak}) and most collisional data are available
  in the BASECOL \citet{basecol} and LAMDA \citet{lamda}
  databases}. Special emphasis was put on molecules specific to the
Herschel Space Observatory, and among them, water. However, complex
organic molecules continue to play a prominent role in the
understanding of proto-stellar evolution as well as being probes for
various interstellar environments.

Among all organic molecules, formaldehyde (H$_2$CO) is especially
abundant, since it is the first stable molecule resulting from the
hydrogenation of the ubiquitous CO molecule \citep{peters2012}. Being
abundant and displaying a large range of both transition frequencies
and energy levels, formaldehyde is a tool of choice to probe the
physical conditions of the gaseous interstellar matter
\citep{maret2004,kama}. By using models that do not suppose Local
Thermodynamical Equilibrium (LTE), it is possible to reach reliable
estimates of the molecular abundance of H$_2$CO together with the
other parameters of the gas, its temperature and density
\citep{cecilia2003,radex}. {\bf The famous formaldehyde ``anomalous''
  absorption was also shown to provide a probe of the ortho-to-para
  ratio of H$_2$ \citep{troscompt2} and a distance-independent tracer
  of the cosmic star formation history \citep{darling12}.}

Several computations of the excitation rates of H$_2$CO have been
proposed in the literature. Excitation by He atoms, being easier to
perform, has been used for long as a model for excitation by H$_2$,
even if it is known for being an underestimation of unknown precision
\citep{green78,green91,sharma2012}. In our group, we computed
\citep{troscompt} the quenching rates of H$_2$CO by H$_2$, but only
for the low levels of excitation of ortho-formaldehyde (only the first
10 levels) and for low kinetic temperatures ($T\lesssim
100\,$K). These computations were based on a high precision Potential
Energy Surface (PES) for the van der Waals interaction $ \mathrm{H_2CO
  - H_2}$. In the present paper, we extend our previous computations
to a much broader range of rotational energies and temperatures, for
both ortho- and para-H$_2$CO, using the same PES. Also, in order to
assess the precision of this PES, we compare experimental pressure
broadening cross sections measured by \citet{delucia} to our own
computations, in a manner similar to our recent works on the
rotational excitation of H$_2$O and CO \citep{drouinPB,jqrst}.

We organize the paper as follows. Section~\ref{sec:sc} describes the
details of the scattering computation. Section~ \ref{sec:res} shows
the results of our computations including cross sections, rates and
pressure broadening. We end (section~\ref{sec:ccl}) with a discussion
and a conclusion.

\section{Scattering calculations}\label{sec:sc}

As mentioned above, the PES for the interaction of H$_2$CO and H$_2$
has been described in our previous paper \citep{troscompt}. The PES
was computed for frozen monomer geometries (rigid-rotor
approximation). The geometries were those of the averaged distances
and angles, at the ground vibrational state for both H$_2$CO and
H$_2$. Using average ground state geometries instead of equilibrium
geometries have been shown to give more reliable results
\citep{valiron08}.  \textbf{Since no modifications were performed on
  the PES, neither on the \textit{ab initio} computations nor on the
  fits,} the reader is deferred to \citet{troscompt} for all necessary
details of the \textit{ab initio} and fitting procedures (see also
\citet{rist2011} for the fitting procedure).

Both H$_2$CO and H$_2$ have each two identical $^1$H nuclei, of
nuclear spin $1/2$. Hence, both exist in para and ortho spin
states. For H$_2$, the para state (total nuclear spin $I=0$) has even
rotational states $J_2=0,2,\ldots$ (We denote by $J_2$, the rotational
quantum number of H$_2$). The ortho state ($I=1$) on the opposite, has
odd rotational states, $J_2=1,3,\ldots$. The H$_2$ rotational constant
is taken at $B=60.853\, \mathrm{cm^{-1}}$. In the PES, the H-H
distance is taken at $R=1.448761$~\AA, its average value at ground
vibrational level \citep{troscompt}. The H$_2$CO molecule is an
asymmetric rotor, with rotational constants (in $\mathrm{cm}^{-1}$) :
$A = 1.29543$, $ B = 1.13419 $, $ C = 9.40552$ and centrifugal
parameters $ D_J = 0.251\,10^{-5}$, $ D_{JK} = 0.431\,10^{-4}$, and
$D_K = 0.648\,10^{-3}$ .  The rotational constant $C$ is along the
electric dipole, on the $C_{2v}$ symmetry axis \footnote{This is the
  \texttt{MOLSCAT} convention, differing from usual spectroscopic
  conventions by the ordering of the axes.}.  Describing the
rotational states by the usual $J_{K_A K_C}$ rotational
(pseudo-)quantum numbers, para states ($I=0$) of H$_2$CO correspond to
$K_A $ even and ortho states ($I=1$) to $K_A$ odd (the main quantum
rotational number for H$_2$CO is denoted by $J$ throughout this
paper).

Scattering calculations were performed for all levels with a
rotational energy $E_{\text{rot}}\leq 210 \,\mathrm{cm^{-1}} $, that
is up to $J_{K_A K_C} = 10_{\,3\,7} $ for ortho-H$_2$CO and $J_{K_A
  K_C}= 7_{\,4\,4}$ for para-H$_2$CO. We computed the inelastic cross
sections with a collision energy $1\leq E_{\text{coll}}\leq 1000
\,\mathrm{cm^{-1}}$. This slightly extends the previous H$_2$CO-He
computations of \citet{green91} and extends our previous H$_2$CO-H$_2$
computations by an order of magnitude.  Experimental \citep{bocquet96}
and computed para and ortho rotational levels of H$_2$CO are given in
table \ref{tab:levels}, along with the level numbering used in the
results. Note that restricting energies to $E\leq 210
\mathrm{\,cm^{-1}}$ entail $K_a\leq 4$. Also, for high enough $J$
values, pairs of successive $K_A\,K_C$ and $K_A\, K_C+1$ levels tend
to be degenerate, for all practical purposes. This degeneracy is all
the more precise that the $K_C$ values are small.  Inspection of
table~\ref{tab:levels} shows that the experimental - theoretical
energy level differences $\Delta E$ are very small: $\left| \Delta
E\right|\leq \mathrm{0.43\, cm^{-1}}$ (mean value $\left<\left|\Delta
E \right|\right> = 0.024 \,\mathrm{cm}^{-1}$).

All scattering calculations have been performed with the OpenMP
version of the {\texttt{MOLSCAT}} code\footnote{Repository at
  http://ipag.osug.fr/$\sim$afaure/molscat/index.html }.  The reduced
mass for H$_2$CO-H$_2$ is 1.889053~amu. The coupled-channel (CC) and
coupled-states (CS) equations were integrated using the diabatic
modified log-derivative propagator.
 
 The rotational basis set is devised as follows. For all scattering
 energies, if $J_{K_AK_C}$ is the last open channels, the $J'=J+1,
 J+2$ values were added to the basis.  However, since a given $J$
 rotational number spans a large amount of rotational energy,
 rotational levels were capped, at
 $E_{\text{max}}=400\mathrm{\,cm^{-1}}$ for collision energies
 $E_{\text{tot}}\leq 175\,\mathrm{cm^{-1}}$, and increasing
 progressively to $E_{\text{max}}=1200\mathrm{\,cm^{-1}}$ for
 $E_{\text{tot}}\sim 1000\,\mathrm{cm^{-1}}$. \textbf{These large
   values of $E_{\text{max}}$ are needed to converge cross sections; a
   similar effect was previously observed for methyl-formate colliding
   with Helium \citep{methyl}}.

 The rotational basis for ortho-H$_2$ is $J_2=1$. It has been shown
 for many systems that including $J_2=3$ in the basis does not have a
 noticeable influence for temperatures as low as 300~K
 (e.g. \citet{daniel2011}). \textbf{The rotational basis for
   para-H$_2$ proved difficult to settle.  We were able to use the
   $J_2=0,2$ basis set for CC para-H$_2$ - ortho-H$_2$CO collisions,
   with $E_{\text{tot}} \leq 130\,\mathrm{cm}^{-1}$
   \citep{troscompt}. For para-H$_2$CO, because of the level
   structure, the $J_2=0, 2$ basis for CC computations proved to be
   practically impossible for $E_{\text{tot}} \gtrsim
   50\,\mathrm{cm}^{-1}$. As a result, we resorted only to a $J_2=0$
   basis, both for ortho- and para-H$_2$CO, stretching the CC
   computations as high as possible and continuing with the CS
   approximation. In order to assess the importance of the $J_2=2$
   channel, however, we complemented the ortho-H$_2$CO data with a
   full CS $J_2=0,2$ computation up to $E_{\text{tot}} \sim
   900\,\mathrm{cm}^{-1}$, using a coarse energy grid. This allowed us
   to check {\it i)} that cross sections for transitions with
   $J_2=2\rightarrow 0$ are negligible, {\it ii)} that cross sections
   for transitions with $J_2=2\rightarrow 2$ are very similar to those
   with $J_2=1\rightarrow 1$ (as observed for other systems, see
   e.g. \citet{daniel2011}) and {\it iii)} that the difference between
   the basis sets $J_2=0$ and $J_2= 0, 2$ \emph{decreases} with
   increasing collision energy, from an average of $\sim 30~\%$ to
   below 10~\%.}

Because of the large number of expansion terms of the potential
function in the spherical harmonic basis \citep{troscompt}, several
strategies have been devised in order to converge the inelastic
scattering computations in a reasonable amount of time (We arbitrarily
tried to limit ourselves to 72 hours of clock time, for one energy
scattering point, on 12 CPU cores).  The radial propagation used a
step size parameter $\textsc{steps} = 15$ except at collision energies
below 10~cm$^{-1}$ where was \textsc{steps} progressively increased up
to 50. Also, \textsc{rmax} values were progressively increased in the
low-energy regime from default to 100. Other propagation parameters
were taken as the \texttt{MOLSCAT} default values.

For total energies above approximately 330~cm$^{-1}$ the coupled
channels (CC) approach, exact in the fully converged limit, proved to
be impractical. We thus had to resort to the usual coupled states (CS)
approximation, with all its shortcomings. Like in our recent HDO-H$_2$
rate calculations \citep{wiesenfeldHDO}, we used an additive constant
to scale appropriately the CS cross sections by their counterpart CC
values. Overlap of the CC and CS calculations showed the validity of
these corrections. \textbf{The statistics of difference $\delta$
  between the CC and CS cross-sections was also examined. We found the
  following values (in $\text{\AA}^2$, $s$, the standard deviation):
  $\left<\left|\delta\right|\right>\leq0.5$, $s(\delta)\leq 2$.} Also,
for collisions with ortho-H$_2$, the total angular momentum
$J_{\text{tot}}$ was not stepped by unit values, as is done
usually. For CC calculation above
$E_{\text{coll}}>200\,\mathrm{cm^{-1}}$, a step in $J_{\text{tot}}$
value $ \textsc{jstep}=4$ was used, with a careful checking of the
convergence of the $\textsc{jstep}>1$ procedure. This procedure was
used throughout the CS computations.

The energy grid was chosen to guarantee a good description of the
resonances, including those pertaining to the high lying rotational
states. Also, since we aim at rates for temperatures up to 300~K, a
particular care was taken to ensure both some economy in the
computational load and a good convergence of the rate computation. Let
us recall that the quenching rate from state $i$ to state $j$,
$k_{j\leftarrow i}(T)$ (in cm$^3\,$sec$^{-1}$) is related to the
inelastic cross section $\sigma_{j\leftarrow i}(E)$ (in \AA$^2$, with
$E$, the collision energy) by the well known Boltzmann average :
\begin{equation}
k_{j\leftarrow i}(T) = \sqrt{\frac{8}{\pi\mu}\frac{1}{T^3}}\,\int_0^\infty \sigma_{j\leftarrow i}(E) E \exp(-E/T) \,
{\mathrm d}E ,\label{eq:rates}
\end{equation}
where $T$ and $E$ are expressed in the same units, and $\mu$ is the
collisional reduced mass.  The probability density $E \exp(-E/T)$
going very slowly down with energy, it is customary, for calculating
$k_{j \leftarrow i}(T=T_0)$, to compute $\sigma_{j\leftarrow i}(E)$ up
to $E\sim 10\, T_0$ (e.g., \citet{dubernet2009}). This approach is
prohibitively time-intensive for a heavy molecule like H$_2$CO, even
within the CS approximation. For high lying rotational states, we
inspected carefully the actual numerical convergence of
equation~\ref{eq:rates} and stopped our energy grid as soon as the
rate was saturated by 10\% in general, and 25\% for the highest
levels, with $E_{up}> 150 \,\mathrm{cm}^{-1}$.

The computation of the pressure broadening cross-sections
\citep{lwPB,drouinPB} necessitates a very fine energy grid and a good
control of the elastic cross sections, much more difficult to obtain
than the corresponding inelastic sections.  Hence, we had to
re-calculate all the $S$-matrices with collision energies from 2 to
$70\, \mathrm{cm^{-1}}$.Then, we compute the $\sigma^{\text{PB}}(T)$
in order to compare it with the experimental results of
\citet{delucia}, in a temperature range of $10 - 30
\,\mathrm{K}$. Both collisions with para-H$_2$ (basis set, $J_2=0,2$)
and ortho-H$_2$ (basis set, $J_2=1$) were performed on an identical
fine energy grid \textbf{and in the CC formalism}.

\begin{table*}
\begin{minipage}{15cm}
\caption{H$_2$CO rotational levels, in cm$^{-1}$. The experimental values are  from \citet{bocquet96}; the 
\textsc{molscat} values are calculated via the rotational constants described in the text.}\label{tab:levels}
\begin{tabular}{lcccll|lcccll}\small
\\
\hline\hline
\multicolumn{6}{c|}{Para H$_2$CO} & \multicolumn{6}{c}{Ortho H$_2$CO} \\
 Level &$J$ & $K_A$ & $K_C$ & \textsc{Molscat} & Experimental & Level&  $J$ & $K_A$ & $K_C$ & \textsc{Molscat} & 
Experimental\\
 \hline 
 1&0 &  0 &  0 &         0.00000  &           0.00000 & 1 & 1 &  1 &  1 &        10.53897  &          10.53900 \\ 
 2&1 &  0 &  1 &         2.42961  &           2.42960 & 2 &1 &  1 &  0 &        10.70021  &          10.70010 \\ 
 3 &2 &  0 &  2 &         7.28640  &           7.28640 & 3 &2 &  1 &  2 &        15.23672  &          15.23690 \\ 
 4 &3 &  0 &  3 &        14.56547  &          14.56550 & 4 &2 &  1 &  1 &        15.72044  &          15.72020 \\ 
 5 &4 &  0 &  4 &        24.25953  &          24.25970 & 5 &3 &  1 &  3 &        22.28172  &          22.28220 \\ 
 6 &5 &  0 &  5 &        36.35891  &          36.35920 & 6 &3 &  1 &  2 &        23.24914  &          23.24870 \\ 
 7 &2 &  2 &  1 &        40.04024  &          40.04020 & 7 &4 &  1 &  4 &        31.67205  &          31.67290 \\ 
 8 &2 &  2 &  0 &        40.04262  &          40.04260 & 8 &4 &  1 &  3 &        33.28429  &          33.28350 \\ 
 9 &3 &  2 &  2 &        47.32780  &          47.32780 & 9 &5 &  1 &  5 &        43.40518  &          43.40670 \\ 
 10 & 3 &  2 &  1 &        47.33970  &          47.33970 &10 & 5 &  1 &  4 &        45.82316  &          45.82200 \\ 
 11 &6 &  0 &  6 &        50.85168  &          50.85240 &11 & 6 &  1 &  6 &        57.47805  &          57.48040 \\ 
 12 &4 &  2 &  3 &        57.04241  &          57.04250 & 12 &6 &  1 &  5 &        60.86224  &          60.86050 \\ 
 13 &4 &  2 &  2 &        57.07810  &          57.07800 & 13 &7 &  1 &  7 &        73.88710  &          73.89070 \\ 
 14 &7 &  0 &  7 &        67.72387  &          67.72520 & 14 &7 &  1 &  6 &        78.39720  &          78.39490 \\ 
 15 &5 &  2 &  4 &        69.18225  &          69.18240 & 15 &3 &  3 &  1 &        88.23817  &          88.23820 \\ 
 16 &5 &  2 &  3 &        69.26541  &          69.26520 & 16 &3 &  3 &  0 &        88.23819  &          88.23830 \\ 
 17 &6 &  2 &  5 &        83.74502  &          83.74530 & 17 &8 &  1 &  8 &        92.62834  &          92.63370 \\ 
 18 &6 &  2 &  4 &        83.91097  &          83.91060 & 18 &4 &  3 &  2 &        97.95762  &          97.95760 \\ 
 19 &8 &  0 &  8 &        86.95982  &          86.96210 & 19 &4 &  3 &  1 &        97.95777  &          97.95780 \\ 
 10 &7 &  2 &  6 &       100.72797  &         100.72860 & 20 &8 &  1 &  7 &        98.42283  &          98.41970 \\ 
 21 & 7 &  2 &  5 &       101.02562  &         101.02500 & 21 &5 &  3 &  3 &       110.10856  &         110.10860 \\ 
 22 & 9 &  0 &  9 &       108.54265  &         108.54640 & 22 &5 &  3 &  2 &       110.10917  &         110.10920 \\ 
 23 &8 &  2 &  7 &       120.12792  &         120.12910 & 23 &9 &  1 &  9 &       113.69738  &         113.70500 \\ 
 24 &8 &  2 &  6 &       120.62140  &         120.62030 & 24 &9 &  1 &  8 &       120.93290  &         120.92900 \\ 
25 &10 &  0 & 10 &       132.45497  &         132.46091 & 25 &6 &  3 &  4 &       124.69196  &         124.69200 \\ 
 26& 9 &  2 &  8 &       141.94120  &         141.94321 & 26 &6 &  3 &  3 &       124.69380  &         124.69380 \\ 
 27 &9 &  2 &  7 &       142.71099  &         142.70931 & 27 &10 &  1 & 10 &       137.08956  &         137.10020 \\ 
 28 &4 &  4 &  0 &       155.16894  &         155.16940 & 28 &7 &  3 &  5 &       141.70886  &         141.70889 \\ 
 29 &4 &  4 &  1 &       155.16894  &         155.16940 & 29 &7 &  3 &  4 &       141.71348  &         141.71350 \\ 
30 &11 &  0 & 11 &       158.67956  &         158.68851 & 30 &10 &  1 &  9 &       145.92017  &         145.91530 \\ 
31 &10 &  2 &  9 &       166.16374  &         166.16690 & 31 &8 &  3 &  6 &       161.16034  &         161.16051 \\ 
32 &10 &  2 &  8 &       167.30709  &         167.30440 & 32 &8 &  3 &  5 &       161.17049  &         161.17059 \\ 
33& 5 &  4 &  1 &       167.31421  &         167.31461 & 33 &11 &  1 & 11 &       162.79999  &         162.81450 \\ 
34& 5 &  4 &  2 &       167.31421  &         167.31461 & 34 &11 &  1 & 10 &       173.37619  &         173.37041 \\ 
 35&6 &  4 &  2 &       181.88960  &         181.88989 & 35 &9 &  3 &  7 &       183.04741  &         183.04781 \\ 
36& 6 &  4 &  3 &       181.88961  &         181.88989 & 36 &9 &  3 &  6 &       183.06768  &         183.06790 \\ 
37&12 &  0 & 12 &       187.20017  &         187.21330 & 37 &12 &  1 & 12 &       190.82361  &         190.84309 \\ 
38 &11 &  2 & 10 &       192.79099  &         192.79581 &38 &12 &  1 & 11 &       203.29127  &         203.28461 \\ 
39 &11 &  2 &  9 &       194.42165  &         194.41750 &39 &10 &  3 &  8 &       207.37096  &         207.37180 \\ 
40&  7 &  4 &  3 &       198.89572  &         198.89600 &40 &10 &  3 &  7 &       207.40854  &         207.40891 \\ 
41 & 7 &  4 &  4 &       198.89574  &         198.89600  & & & & & & \\
 
\hline
\end{tabular}
%% Any table notes must follow the \end{tabular} command.
\end{minipage}
\end{table*}

\section{Results and discussion}\label{sec:res}

\subsection{Cross sections}\label{sec:sections}

The inelastic cross sections have a general shape that is similar to
all earlier findings, for collisions of a molecule with H$_2$.  As
usual, the inelastic scattering with $J_2=0$ may be markedly different
from the scattering with $J_2>0$. This was observed and thoroughly
discussed for H$_2$O and HDO scattering computations
\citep{dubernet2009,daniel2011,wiesenfeldHDO,faureHDO}, and observed
for a wide range of other collisions, SO$_2$ and Cl atoms being recent
examples \citep{so2,cl}.  Experiments with H$_2$ colliding with water
molecules also extensively confirm this difference
\citep{drouinPB,yangDCS}.  The situation with formaldehyde colliding
with H$_2$, $J_2=0$ and $J_2>0$ is however less clear, as some
ortho-H$_2$ and para-H$_2$ collisions are nearly identical, especially
for small sections and large $\Delta J$.  Examples of astrophysical
significative cross-sections $\sigma(E_{\text{coll}})$ are given in
figure~ \ref{fig:three_sections}. Those three cases are representative
of all these astrophysically relevant types of sections that we
examined: $\sigma(E_{\text{tot}};\,J_2=0) \ll
\sigma(E_{\text{tot}};\,J_2>0) $ even if the
$\sigma(E_{\text{tot}};\,J_2=0) $ displays a richer resonance
structure, because of the disappearance of the supplementary quantum
coupling $\mathbf{J} +\mathbf J_2= \mathbf J_{12}$. In each case, we
observe that $\sigma(E_{\text{coll}};\,J_2=1) \simeq
\sigma(E_{\text{coll}};\,J_2=2) $, in structure and
magnitude. Remember, however, that the threshold for scattering with
$J_2=2$ is $B(H_2)\times 4 = 243.4 \mathrm{\,cm}^{-1}$ higher than for
$J_2=1$. A full and reliable computation of H$_2$CO scattering with
para-H$_2$, $J_2\geq 2$ is thus impossible for all practical purposes,
in the present computer configurations.

%\clearpage
\subsection{Rates}\label{sec:rates}

All quenching rates for all levels of table~\ref{tab:levels} are
computed for the same temperature grid as \citet{green91},
$10\,\mathrm{K} \leq T \leq 300\,\mathrm K$.  The full table is
deposited in the LAMDA database \citep{lamda} and BASECOL database
\citep{basecol}, and may be asked to the authors. The rates with
ortho-H$_2$ are based on the $J_2=1$ sections only. In all cases,
rates with para-H$_2$ are given as a Boltzman average over the
populations of $J_2=0,2$, with the further approximation of
$\sigma(E_{\text{coll}}; J_2=2)\simeq \sigma(E_{\text{coll}}; J_2=1)$,
when necessary (see above).  The influence of the $J_2=2$ initial
states may indeed be very large, as was also observed, in another
context, for the pressure broadening of H$_2$O by H$_2$
\citep{drouinPB}.
 
 Figure \ref{fig:compare_rates} compares globally present critical
 densities and critical densities from \citet{green91}, for
 electric-dipole allowed transitions. We compare our present rates
 with para-H$_2$ with the properly scaled rates with He. We have the
 following definition on the critical density $n^*$:
\begin{equation}
n^*_i=\frac{\sum_{j'<i} A_{j'\leftarrow i}}{\sum_{j<i} k_{j\leftarrow i}(T)}\label{eq:crit}
\end{equation}
where $i$ is the level under scrutiny and $j'$ denotes all levels
connected by a radiative transition, while $j$ spans all levels. Left
panel shows the scatter of critical densities ratios at
$T=300\,\mathrm K$, while on the right panel, all ratios are averaged
and plotted against temperature. The scatter is moderate, with no
ratio exceeding 3. The right panels shows the evolution of the average
$\left<n^*_i\right>$ as a function of $T$. For higher temperatures,
the effect of H$_2$ being different of He diminishes, since higher
collisions energies are more sensitive to the hard walls of the
target, where the potential grows exponentially (repulsion of the wave
functions). H$_2$ and He become more similar, and the rates tend one
towards the other.
 
\subsection{Pressure broadening}

In order to assess the reliability of the PES, we found it useful to
compare measured and computed pressure-broadening cross sections,
$\sigma^{\mathrm{PB}}(T)$ \citep{lwPB,drouinPB,jqrst}. The main
advantage of pressure broadening is that experiments and computations
held both absolute quantities, with no scaling involved, rending the
comparison very meaningful. Unfortunately, only a very limited set of
$ \mathrm{H_2CO}$ pressure broadening data exists for an H$_2$ buffer
gas \citep{delucia}, for very low temperatures. The experimental
results for the $2_{12}\rightarrow 3_{13}$ transition are depicted in
figure \ref{fig:broadening}, along with full CC calculations of
$\sigma^{\mathrm{PB}}(T)$. We see that the para-H$_2$ and ortho-H$_2$
computation bracket the experimental values, which are very well
simulated by an ortho- to para-H$_2$ ratio (\textsc{opr})
corresponding to a pseudo-equilibrium at $\sim$ 50~K,\textbf{
  corresponding to a OPR of 0.27}.
That the \textsc{opr} of H$_2$ may vary during the collisional cooling
experiment has been proved with cell walls covered with amorphous
water \citep{drouinPB}. Nothing is known for formaldehyde, and
discussion with the authors of \citet{delucia} could not settle the
case.  We remain thus with a good plausibility argument, as long as
the pressure broadening experiments with H$_2$ are not fully
characterized.

\section{Discussion - Conclusion} \label{sec:ccl}

It is important to know which errors are to be expected, and to have
some clues on how these errors might influence astrophysical modeling.
We expect the error on the rates $k(T)$ to be uniformly increasing
from low lying levels to higher lying ones, and also from low
temperature to high temperature. Quantifying this error is very risky,
as errors may arise from all the phases of the rate computation:
\textsl{ab initio} computation, fits and long distance behavior of the
PES, convergence of the CC/CS procedures, convergence of the averaging
procedure of the sections with collision energy.

 Internal consistency with our earlier approach \citep{troscompt} show
 differences less than 10~\%, for $J \leq 5$, $T \leq 50\,\mathrm
 K$. This shows that the convergence error in this domain cannot
 exceed 20-30\%. \textbf{In the high temperature domain ($T \geq 150$
   K), it is safe to assume errors much larger than 30\%, because of
   the poor convergence of formula (\ref{eq:rates}), but still within
   a factor~2. The magnitude of the $\delta$ values as well as the
   similarities between the $J_2=1\rightarrow 1$ and $J_2=2
   \rightarrow 2$ plead in favor of such a conservative value, as does
   the convergence of critical densities between our work and the
   previous collisions with He by \citet{green91}}.

Accuracy of the PES is very difficult to assess, without any firm
experimental comparison, like has been done for H$_2$O and to a lesser
extent CO \citep{yangDCS,drouinPB,jqrst,bordeauxprl}.  Our results
are, however, compatible with measurements of \citet{delucia}.
Because of the great importance and ubiquity of the formaldehyde
molecule, further experiments would give indications on the precision
of the PES and convergence procedures used in this paper.
 
The relevance of the ortho-to-para ratio of H$_2$ has been stressed
several times already, as it may be of crucial importance in order to
correctly model the astrophysical environments. While the difference
in rates for ortho-H$_2$ and para-H$_2$ is large for low lying levels
at low energies, see \citet{troscompt2} for an application, this
difference decreases at larger transition energies, as the hard walls
of the PES play a more important role than the long range
behavior. Indeed, the main increase in $\sigma(J_2>0)$ with respect to
$\sigma(J_2=0)$ is due to the averaging out of the quadrupolar moment
and dipolar polarizability of H$_2$ in its ground rotational
states. The same is true, up to a global scale, for the difference in
behavior between He and ortho-H$_2$.

Extension of these computations to higher levels and higher
temperatures is by no means a difficult task, on the physics point of
view, because of the rigid rotor structure of H$_2$CO, its first
bending frequency (the out-of-plane bend) arising at 1167.3~cm$^{-1}$
\citep{annrev}. The true limiting factor arises from numerical load,
with very large $N\times N$ matrices to invert and propagate ($N >
2000$).  The same situation arises for the excitation of heavier
complex organic molecules, like methyl formate or dimethyl ether,
which display many spectral lines very far from LTE, in various
spectral surveys like \citet{TIMASSS}.  Unfortunately, quasi classical
trajectories methods are limited for all those cases because of
ambiguities arising in the subsequent quantization of rotational
levels \citep{ambiguities}. Use of very large grid of computers and
combinations of OpenMP/MPI approaches may overcome these difficulties.

We have calculated an extensive set of low to medium temperature
quenching rates, for all levels of H$_2$CO below
$210\,\mathrm{cm^{-1}}$. These rates are ready to be incorporated in
the various non-LTE models for the interstellar medium.  At low
temperatures, differences with earlier rates of H$_2$CO colliding with
Helium are very important and they remain noticeable at all
temperatures, with ratios up to 50\% at 300~K, where they are the most
similar. {\bf Since \citet{troscompt} was tailored to be very precise
  at low temperatures for ortho-H$_2$CO, we still recommend to use
  those rates for applications at $T\leq 30$~K. The present rates
  should have a large importance on the inferring of H$_2$CO column
  densities, away from LTE conditions.

\section*{acknowledgments}

Many discussions with C. Ceccarelli and members of the CHESS team are
gratefully acknowledged.  The authors thank generous funding through
the ANR and the CNES agencies, thanks to the FORCOMS contract
(ANR-08-BLAN-022) and the CHESS Herschel Space Observatory Key Program
funding.  This work was also supported by the CNRS-INSU national
program ``Physique et Chimie du Milieu Interstellaire''.  All
calculations presented in this paper were performed at the ``Service
Commun de Calcul Intensif de l'Observatoire de Grenoble (SCCI)''.

\begin{figure*}
\vbox {\vfil  \includegraphics[width=17cm]{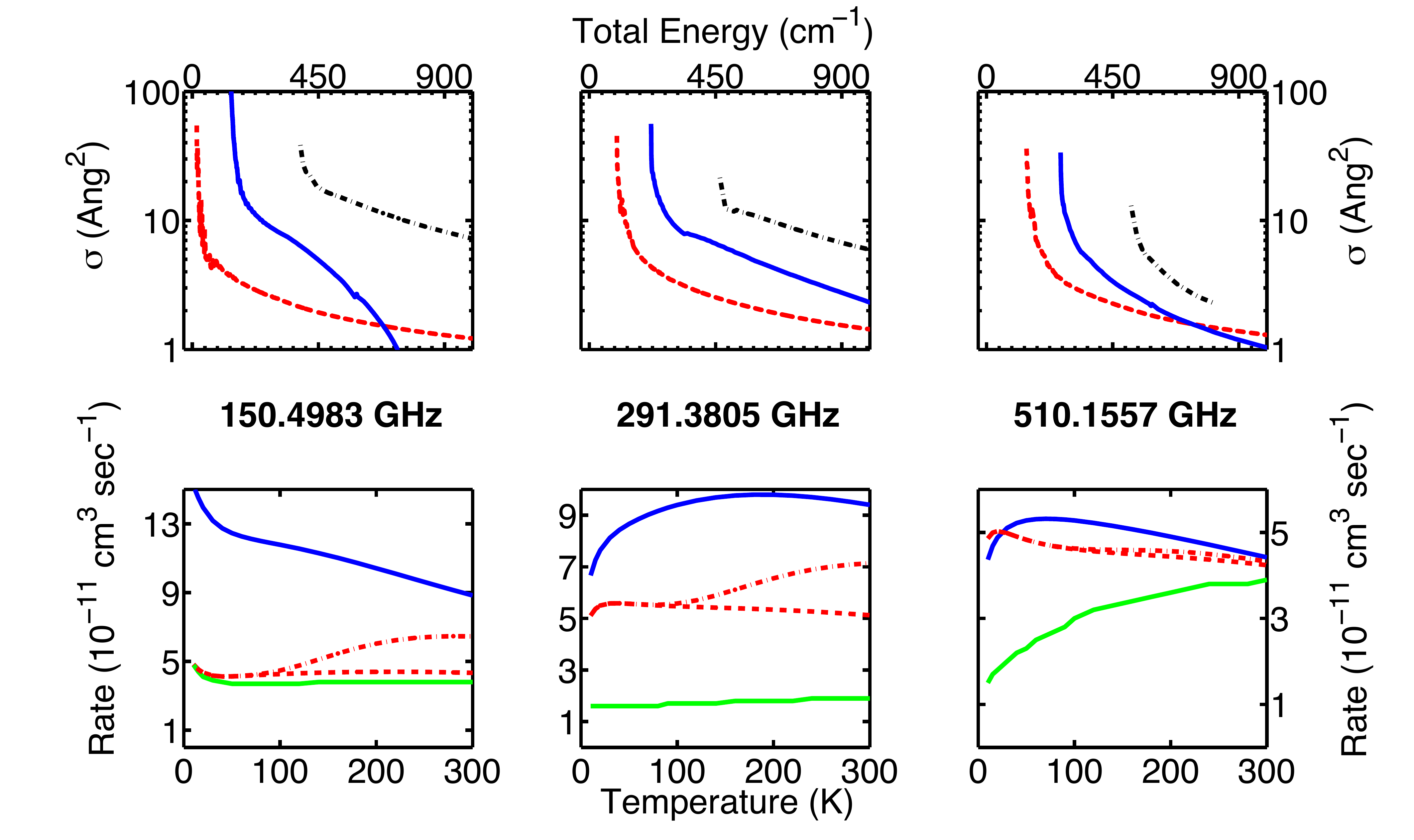}
\caption{(Color online) \textbf{Cross-sections (upper three panels)
    and rates (lower three panels) for three widely observed quenching
    transitions of ortho-H$_2$CO (first column : $2_{11}\rightarrow
    1_{10}$, at $E_{up}=22.62\,$K; second column : $4_{32}\rightarrow
    3_{31}$, at $E_{up}=140.93\,$K; third column : $7_{35}\rightarrow
    6_{34}$, at $E_{up}=203.9\,$K). In all panels the red (dashed)
    line denotes the H$_2$ rotational $J_2=0$ initial and final state,
    the blue (solid) line, the same with $J_2=1$ and the black
    (dashed-dotted), the same with $J_2=2$. The red dash-dotted rates
    include the influence of the population of the $J_2=2$ para state
    of H$_2$ (see text).} Rates and sections with $J_2=0
  \leftrightarrow J_2=2$ are two or three orders of magnitude lower
  and not depicted. Rates for collisions with Helium \citep{green91}
  are the lowest rates in all three inferior panels (green color
  online). Note that for cross sections, for sake of clarity, the
  energy is the \emph{total} energy. Note also that cross sections
  display usual resonance patterns which are not conspicuous at this
  log scale.} \vfil }\label{fig:three_sections}
\end{figure*}
\clearpage

\begin{figure*}
\vbox {\vfil
\includegraphics[width=15cm]{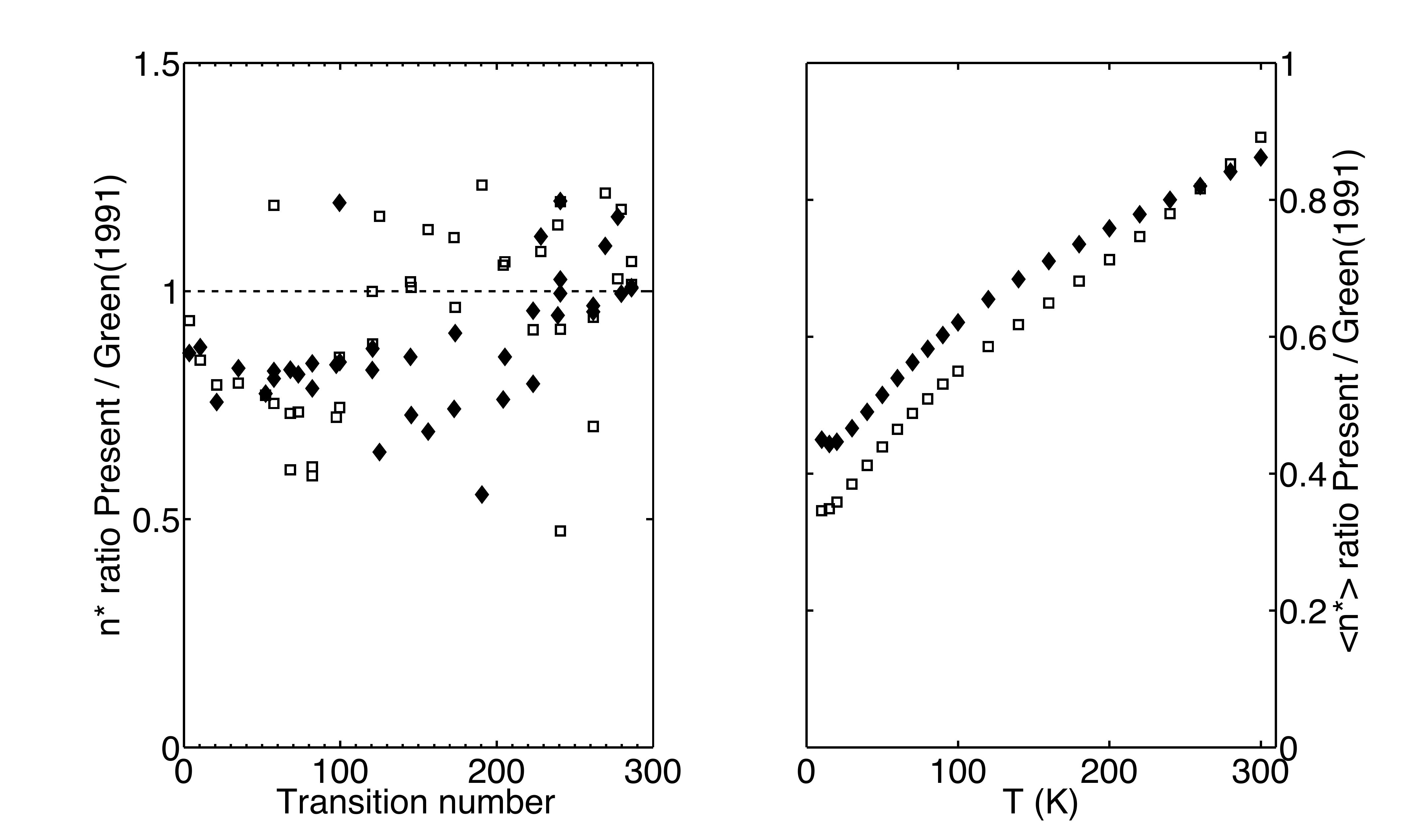}
\caption{ Left panel: ratio of critical densities (eq. \ref{eq:crit}),
  present work to \citet{green91} properly scaled, for all levels, at
  $T=300\,\mathrm K$. Right panel, average over all critical densities
  ratios, as a function of temperature. Both panels : open squares,
  ortho-H$_2$CO with para-H$_2$ ($J_2=0$) or He; filled diamonds
  para-H$_2$CO with para-H$_2$ ($J_2=0$) or He. }\vfil}
\label{fig:compare_rates}
\end{figure*}
\clearpage
\begin{figure}
%\epsscale{.80}
\includegraphics[width=0.4\textwidth]{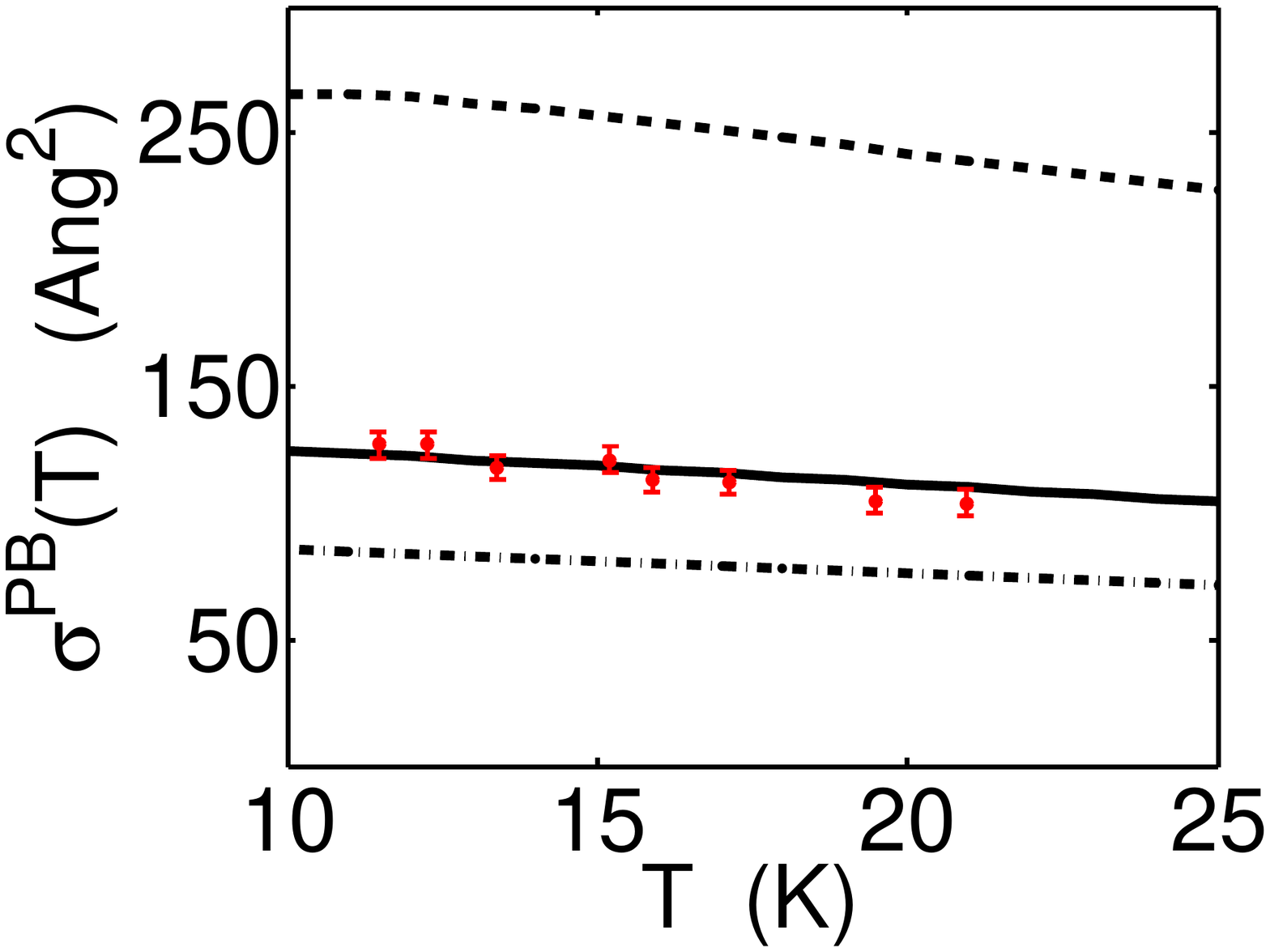}
\caption{(Color online) Pressure broadening of the $2_{12}\rightarrow
  3_{13}$ transition. Red symbols, measurements of
  \citet{delucia}. Dashed and dot-dashed lines, present computations
  with respectively ortho-H$_2$ and para-H$_2$. Full line,
  ortho-to-para of H$_2$ equivalent to a spin temperature of
  50~K.}\label{fig:broadening}
\end{figure}

\end{document}